\documentclass[12pt]{article}
\usepackage{graphicx}

\def\ignore#1{{}}

\let\oldtheequation=\theequation
\def\doteqs#1{\setcounter{equation}{0}            
\def\theequation{{#1}.\oldtheequation}}
\newcounter{sxn}
\def\sx#1{\addtocounter{sxn}{1} \vskip 1.cm  \goodbreak
\noindent{\large\bf\leftline{\thesxn.~~#1}} \nobreak \vskip -.5cm}
\def\sxn#1{\sx{#1} \doteqs{\thesxn}}

\newcounter{axn}

\date{}

\newdimen\mybaselineskip
\newcommand{\beeq}{\begin{equation}}
\newcommand{\eneq}{\end{equation}}
\newcommand{\beqn}{\begin{eqnarray}}
\newcommand{\eeqn}{\end{eqnarray}}

\def\la{\raise.16ex\hbox{$\langle$}\lower.16ex\hbox{}  }
\def\ra{\, \raise.16ex\hbox{$\rangle$}\lower.16ex\hbox{} }

\def\psibar{ \psi \kern-.65em\raise.6em\hbox{$-$} \lower.6em\hbox{} }
\def\psibarb{ \psi \kern-.65em\raise.6em\hbox{$-$}  }

\begin{document}

\thispagestyle{empty}

\baselineskip=12pt

%{\small \noindent \mydate  \hfill UW}

%{\small \noindent Ramin\hfill   }

\vspace*{3.cm}

\begin{center}  
{\LARGE \bf  The Highly Damped Quasinormal Modes of $d$-Dimensional Reissner-Nordstr$\ddot{\rm o}$m Black Holes in the Small Charge Limit}
\end{center}

\baselineskip=14pt

\vspace{3cm}
\begin{center}
{\bf  Ramin G. Daghigh, Gabor Kunstatter, Dave Ostapchuk, and Vince Bagnulo}
\end{center}

\centerline{\small \it Physics Department, University of Winnipeg, Winnipeg, Manitoba, Canada R3B 2E9}
\centerline{\small \it and}
\centerline{\small \it Winnipeg Institute for Theoretical Physics, Winnipeg, Manitoba}
\vskip 1 cm
\centerline{} 

\vspace{1cm}
\begin{abstract}
We analyze in detail the highly damped quasinormal modes of $d$-dimensional Reissner-Nordstr$\ddot{\rm{o}}$m black holes with small charge, paying particular attention to the large but finite damping limit in which the Schwarzschild results should be valid. In the infinite damping limit, we confirm using different methods the results obtained previously in the literature for higher dimensional Reissner-Nordstr$\ddot{\rm{o}}$m black holes. Using a combination of analytic and numerical techniques we also calculate the transition of the real part of the quasinormal mode frequency from the Reissner-Nordstr$\ddot{\rm{o}}$m value for very large damping to the Schwarzschild value of $\ln(3) T_{bh}$ for intermediate damping. The real frequency does not interpolate smoothly between the two values. Instead there is a critical value of the damping at which the topology of the Stokes/anti-Stokes lines change, and the real part of the quasinormal mode frequency dips to zero. 
\baselineskip=20pt plus 1pt minus 1pt
\end{abstract}

%%%%%%%%%%% 1 %%%%%%%%%%
\newpage

\sxn{Introduction}

\vskip 1cm
There exist by now many calculations of the highly damped quasinormal modes (QNMs) of a large variety of black holes. Since the highly damped modes are not directly observable, these calculations are mainly motivated by the proposal, originating with Hod\cite{Hod} and more recently propounded by Dreyer\cite{Dreyer} in the context of loop quantum gravity, that these modes may be providing information about the semi-classical quantum spectrum of black holes.  This proposal was prompted in part by the apparent universality of the real part of the frequency in the highly damped limit, as well as by its special numerical value for Schwarzschild black holes,
namely:
\beeq
\omega_R \mathop{\longrightarrow}_{|\omega_I|\to\infty}\ln(3)T_{bh}/\hbar ~,
\eneq
where $T_{bh}$ is the Hawking temperature of the black hole. The frequency and the damping are proportional to the temperature on dimensional grounds for single horizon black holes, but the coefficient $\ln(3)$ allows, on application of the Bohr correspondence principle, an elegant statistical interpretation\cite{Hod, Bekenstein-1} of the resulting Bekenstein-Hawking entropy spectrum. We refer the reader to the literature for details\cite{Natario}. Although this special value has been shown to be valid for a large class of single horizon black holes\cite{Motl2,Gabor3,Das1,Ramin1} this simple relationship between frequency and temperature  breaks down for multi-horizon black holes.

It was shown first by Motl and Neitzke\cite{Motl2} and then confirmed by Andersson and Howls\cite{Andersson} that for Reissner-Nordstr$\ddot{\rm{o}}$m (R-N) black holes the real part of the highly damped QNM frequency in four dimension approaches $\ln(5)T_{bh}$ as the charge goes to zero. The general validity of this result was verified by Natario and Schiappa\cite{Natario-S}, who calculated the R-N highly damped spectrum in arbitrary spacetime dimensions. The apparent contradiction between the zero charge limit of the R-N case, and the Schwarzschild value was explained heuristically by the authors of \cite{Andersson}. They noted, that, while the $\ln(5)$ represented the correct R-N result for very high damping, one expects an intermediate range of damping for which the Schwarzschild value of the real frequency is correct. Order of magnitude arguments suggest this range in 4-$d$ to be:
\beeq
1<<|\omega|^2M^2 << M^8/q^8
\eneq

The purpose of the present paper is two-fold. First, we calculate
 the highly damped QNMs of $d$-dimensional R-N black holes with the small charge using the techniques of Andersson and Howls\cite{Andersson}, thereby confirming the results of Natario and Schiappa\cite{Natario-S}, who used the monodromy method of Ref.\cite{Motl2,Motl1}.  Secondly, and more significantly, we use a combination of analytic and numerical techniques to analyze the limit of large but finite damping where the Schwarzschild limit is approached. For four and five spacetime dimension R-N black holes we explicitly calculate the spectrum in the transition region from $\ln(3+4\cos({d-3 \over 2d-5}\pi))T_{bh}$ for very large damping to the Schwarzschild value of $\ln(3) T_{bh}$. The real frequency does not interpolate smoothly between the two values. Instead there is generically a critical value of the damping at which the Stoke/anti-Stokes lines change topology and the real part of the frequency dips to zero. This behaviour seems to be the analogue of the algebraically special frequencies (for a recent discussion see Berti\cite{Berti1}) that mark the onset of the high damping regime. We provide general arguments that suggest the persistence of this behaviour for all spacetime dimensions.

The paper is organized as follows. In Section 2 we describe the general formalism. Section 3 summarizes the results of the calculation in the large damping limit for metric perturbations (tensor, vector, and scalar) of Schwarzschild and R-N.  The former reviews well known results while the latter provides independent confirmation of \cite{Natario-S}. Section 4 presents the details of the calculation for axial perturbations in the intermediate region in four dimensions, while Section 5 does the same for five dimensions. (Tensor and scalar perturbations cannot be done using our numerical techniques, because a crucial term in the effective potential vanishes precisely in that limit.) The final section describes general arguments for the persistence of the qualitative features of our results in higher dimensions, and closes with some conclusions.

\sxn{General Formalism}

It was shown by Ishibashi and Kodama in \cite{Ishibashi1}, \cite{Ishibashi2}, and \cite{Ishibashi3} that various classes of non-rotating black hole metric perturbations in a spacetime with dimension $d>3$ are governed generically by a Schr$\ddot{\mbox o}$dinger wave-like equation of the form:
\beeq
{d^2\psi \over dz^2}+\left[ \omega^2-V(r) \right]\psi =0 ~,
\label{Schrodinger}
\eneq
for perturbations that depend on time as $e^{-i\omega t}$.  The Tortoise coordinate $z$ is defined by 
\beeq
dz ={dr \over f(r) }~,
\label{tortoise}
\eneq
where $f(r)$ is related to the spacetime geometry, and is given by 
\beeq
f(r)=1-{2\mu \over r^{d-3}}+{\theta^2 \over r^{2d-6}}-\lambda r^2~.
\label{function f}
\eneq
The ADM mass, $M$, of the black hole is related to the parameter $\mu$ by
\beeq
M = {(d-2)A_{d-2} \over 8 \pi G_d}\mu~,
\eneq
where $A_n$ is the area of a unit $n$-sphere,
\beeq
A_n={2\pi^{n+1 \over 2} \over \Gamma\left({n+1 \over 2}\right) }~.
\eneq
The electric charge, $q$, of the black hole is,
\beeq
q^2 = { (d-2)(d-3)\over 8 \pi G_d }\theta^2~,
\eneq
while the value of the cosmological constant, $\Lambda$, is given by 
\beeq
\Lambda = { (d-1)(d-2)\over 2 }\lambda~.
\eneq
In this paper we only deal with asymptotically flat black holes where $\lambda=\Lambda=0$.  The units we will be using are $c=G_d=1$.

The effective potential $V(r)$, was found explicitly by Ishibashi and Kodama\cite{Ishibashi1,Ishibashi2,Ishibashi3} for scalar (reducing to polar at $d=4$), vector (reducing to axial at $d=4$), and tensor (non-existing at $d=4$) perturbations.  The effective potential for tensor perturbations has been shown \cite{Gibbons, Konoplya} to be equivalent to that of the decay of test scalar field in a black hole background.  The effective potential is zero at both the horizon ($z\rightarrow -\infty$) and spatial infinity ($z\rightarrow \infty$).  In the case of QNMs, the asymptotic behavior of the solutions is chosen to be
\beeq
\psi(z) \approx \left\{ \begin{array}{ll}
                   e^{-i\omega z}  & \mbox{as $z \rightarrow -\infty$ $(x\rightarrow x_h)$}~,\\
                   e^{+i\omega z}  & \mbox{as $z\rightarrow \infty$ $(x\rightarrow \infty)$~,}
                   \end{array}
           \right.        
\label{asymptotic}
\eneq 
which represents an outgoing wave at infinity and an ingoing wave at the horizon.  Since the tortoise coordinate is multi-valued, it is more convenient to work in the complex $r$-plane.  After rescaling the wavefunction $\psi=\Psi/\sqrt{f}$ we obtain
\beeq
\frac{d^2\Psi}{dr^2}+R(r)\Psi=0 ~,
\label{Schrodinger-r}
\eneq 
where
\beeq
R(r)= {\omega^2\over f^2(r)}-U(r)~,
\label{Rr}
\eneq
with
\beeq
U(r)={V(r)\over f^2}+{1\over 2}\frac{f''}{f}-\frac{1}{4}\left(\frac{f'}{f}\right)^2 ~.
\label{Ur}
\eneq
Here prime denotes differentiation with respect to $r$.

\sxn{Large Damping Limit and Universality}

We now consider the QNMs in the infinite damping limit where
\beeq
|\omega^2|\to [{\rm Im}~\omega]^2 \to \infty~.
\eneq
Since we will be using complex analytic techniques, in principle the behaviour of $U(r)$ on the entire complex plane may be relevant.
However, in the infinite limit case, the $\omega^2/f^2$ term in $R(r)$ will dominate $U(r)$ everywhere, unless one of the terms in $U(r)$ diverges. This can only happen at the origin, or at the horizon. However, since $\omega^2/f^2$ also diverges at the horizon, it will dominate there as well in the large damping limit. Thus, only the dominant term of  $U(r)$ near the origin is relevant in this limit and $R(r)$ can be approximated on the entire complex plane by
\beeq
R(r)\sim \frac{\omega^2}{f^2}-\frac{J^2-1}{4r^2}~,
\label{Rr2}
\eneq
where $J$ is determined by the type of metric perturbation.  In particular, for different perturbations we find:
\beeq
J^2-1= \left\{ \begin{array}{ll}
                   -1~~~ \mbox{Tensor Perturbation ($\theta=0$ and $n>2$)}~,\\
                   n(n-2)~~~ \mbox{Tensor Perturbation ($\theta \neq 0$ and $n>2$)}~,\\
                   4n^2-1~~~ \mbox{Vector Perturbation ($\theta = 0$)}~,\\
                   3n(3n-2)~~~ \mbox{Vector Perturbation ($\theta\neq 0$)}~,\\
                   -1~~~ \mbox{Scalar Perturbation ($\theta= 0$)}~,\\
                   {n}(n-2)~~~ \mbox{Scalar Perturbations ($\theta \neq 0$)}~,\\
                   \end{array}
           \right.        
\label{J}
\eneq
where $n=d-2$. The universality of the infinite damping QNM frequency can, to a large extent, be attributed to the simple form of (\ref{Rr2}) in this limit.

In the WKB analysis, the two solutions to Eq. (\ref{Schrodinger-r}) are
approximated by
\beeq
\left\{ \begin{array}{ll}
                   \Psi_1^{(t)}(x)=Q^{-1/2}(x)\exp \left[+i\int_{t}^xQ(x')dx'\right]~,\\
                   \\
                   \Psi_2^{(t)}(x)=Q^{-1/2}(x)\exp \left[-i\int_{t}^xQ(x')dx'\right]~,
                   \end{array}
           \right.        
\label{WKB}
\eneq
where
\beeq
Q^2(r)=R(r)-\frac{1}{4r^2}\sim \frac{\omega^2}{f^2}-\frac{J^2}{4r^2}~
\label{Q^2-general}
\eneq
is shifted by $1/(4r^2)$ in order to guarantee the correct behaviour of the WKB solution at the origin.  In Eq. (\ref{WKB}), $t$ is the simple zero of $Q^2$.

Given the function $Q$, it is possible to determine the WKB condition for asymptotic QNMs. The methodology that we adopt has been explained in details in \cite{Andersson}. First, one determines the zeros and poles of the function $Q$ and consequently the behavior of the Stokes and anti-Stokes lines in the complex $r$-plane.  Stokes lines are the lines on which the WKB phase ($\int Q dr$) is purely imaginary and anti-Stokes lines are the lines on which the WKB phase is purely real.  It turns out that for Schwarzschild black holes in any spacetime dimension, there are two unbounded anti-Stokes lines which extend to infinity on either side of a bounded anti-Stokes line that encircles the event horizon.  In the case of R-N black hole, which has two horizons,  the unbounded anti-Stokes lines are again on either side of a bounded anti-Stokes line encircling both the event horizon and the inner Cauchy horizon. Inside this loop a second bounded anti-Stokes line encircles the inner Cauchy horizon.  The structure of anti-Stokes lines for Schwarzschild and R-N black holes in four and higher dimensions can be found in \cite{Natario-S}.  

To determine the WKB condition on the QNM frequency we start on an unbounded anti-Stokes line where the solution is known due to the boundary condition at infinity:
\beeq
\psi_{initial}=\Psi_{\infty}~.
\eneq 
It is relatively simple to determine how the solutions change as we move along an anti-Stokes line.  Moreover, there are rules that specify how the solution changes when one moves from one anti-Stokes line to another while crossing Stokes lines. The procedure is to move along a contour that goes from the known solution at infinity along the unbounded anti-Stokes line, changes to the bounded anti-Stokes line(s), loops around the event horizon and finally returns to infinity along the same unbounded anti-Stokes line that it started from. This yields a final wave solution $\psi_{final}$ that is related to the one we started from as follows:
\beeq
\psi_{final}=\chi \Psi_{initial}~,
\eneq
where $\chi$ is the monodromy of this contour. Since this contour circles only one pole, namely the horizon, the  monodromy of this contour must be the same as the monodromy of a contour in the vicinity of the pole at the event horizon. This latter monodromy is determined by the boundary condition at the horizon, so that we are left with a consistency condition that determines asymptotic QNMs.
 
Applying the above method to Schwarzschild black holes in four and higher dimensions one arrives at the condition
\beeq
e^{2i\Gamma}=-1-2\cos\left({J\over n}\pi\right)~,
\label{Sch-WKB}
\eneq 
where the appropriate value of $J$ is given in (\ref{J}). $\Gamma$ is the integral of the function $Q$ along a contour encircling the pole at the event horizon in the negative direction.  Using the residue theorem to evaluate $\Gamma$ yields
\beeq
\Gamma=\oint Q dy= -2\pi i \mathop{Res}_{r=|r_h|} Q= -2\pi i\left[{\frac{1}{n-1}(2\mu)^{1/(n-1)}\omega}\right] ~.
\label{Gamma}
\eneq
We can now combine Eqs. (\ref{J}), (\ref{Sch-WKB}), and (\ref{Gamma}) to evaluate $\omega$:
\beeq
{\frac{4\pi}{n-1}(2\mu)^{1\over n-1}\omega} = \ln(3)+(2k+1)\pi i~~~~\mbox{as $k\rightarrow \infty$}
\label{omega_final}
\eneq
This result is valid for all types of metric perturbations in four and higher dimensions.

In the R-N case where $\theta \neq 0$, the same procedure yields the following general WKB condition for four and higher dimensions:
\beeq
e^{2i\Gamma_e}=1-2\left[1+\cos\left({J\over 2n-1}\pi\right)\right](1+e^{-2i\Gamma_i})~,
\label{RN-WKB}
\eneq 
where $\Gamma_e$ and $\Gamma_i$ are the integral of the function $Q$ along a contour encircling the pole at the event horizon and the inner Cauchy horizon respectively in the negative direction.  We can also evaluate these integrals using the residue theorem, and the results are
\beeq
\Gamma_e=-2\pi i \mathop{Res}_{r=|r_+|} Q= -\pi i{\frac{r_+^n \omega}{(n-1)\sqrt{\mu^2-\theta^2}}} ~,
\label{Gamma-e}
\eneq
and 
\beeq
\Gamma_i=-2\pi i \mathop{Res}_{r=|r_-|} Q= \pi i{\frac{r_-^n \omega}{(n-1)\sqrt{\mu^2-\theta^2}}} ~.
\label{Gamma-i}
\eneq
Here $r_+$ and $r_-$ are the locations of the event horizon and the inner Cauchy horizon respectively, which are given by
\beeq
r_\pm=\left(\mu \pm \sqrt{\mu^2- \theta^2}\right)^{1\over n-1}~.
\eneq
Since in this paper we are interested in the Schwarzschild limit of Eq. (\ref{WKB-RN-small q}), let us find $\omega$ in the limit where $\theta\rightarrow 0$, we note that in this limit:
\beeq
\Gamma_e \approx -{2\pi i \over n-1} \omega (2\mu)^{1 \over n-1}  ~,
\eneq 
and
\beeq
\Gamma_i \approx {\pi i \over n-1} \omega \mu^{1 \over n-1} \left( {\theta^2 \over 2\mu^2} \right)^{n \over n-1}~.
\eneq 
Equation (\ref{RN-WKB}) then  takes the simple form
\beeq
\exp\left({4\pi  \over n-1} (2\mu)^{1 \over n-1} \omega \right) \approx -3-4\cos\left({J\over 2n-1}\pi\right)~.
\label{WKB-RN-small q}
\eneq 
The constant $J$ for tensor and scalar perturbations is $n-1$, while for vector perturbations it is $2n$.  Even though  $J$ is different in vector perturbations, but for these perturbations we can write
\beeq
{J_{vector} \over 2n-1}={2n\over 2n-1}=2-{n-1 \over 2n-1}~,
\eneq 
which produces the same result in Eq. (\ref{WKB-RN-small q}) as tensor and scalar perturbations for which 
\beeq
{J \over 2n-1}={n-1 \over 2n-1}~.
\eneq 
Thus, in the small charge limit, the R-N black hole QNM frequency is:
\beeq
{4\pi  \over n-1} (2\mu)^{1 \over n-1} \omega  \approx \ln\left[3+4\cos\left({n-1\over 2n-1}\pi\right)\right] +(2k+1)\pi i~~~~\mbox{as $k\rightarrow \infty$}~.
\label{omega-RN-small q}
\eneq 
Clearly this does not correspond with the Schwarzschild result in Eq. (\ref{omega_final}). The difference, as pointed out in \cite{Andersson} is due to the order in which the limits  $|\omega|\rightarrow \infty$ and $\theta \rightarrow 0$ are taken.  This issue will be addressed in detail in the following sections.

Note that the results in this section are consistent with the results obtained by Natario and Schiappa in \cite{Natario-S}, where the reader can find a detailed calculation of the highly damped QNMs based on the different methodology first used in \cite{Motl2} and \cite{Motl1}.

\begin{figure}[tb]
\begin{center}
\includegraphics[height=10cm]{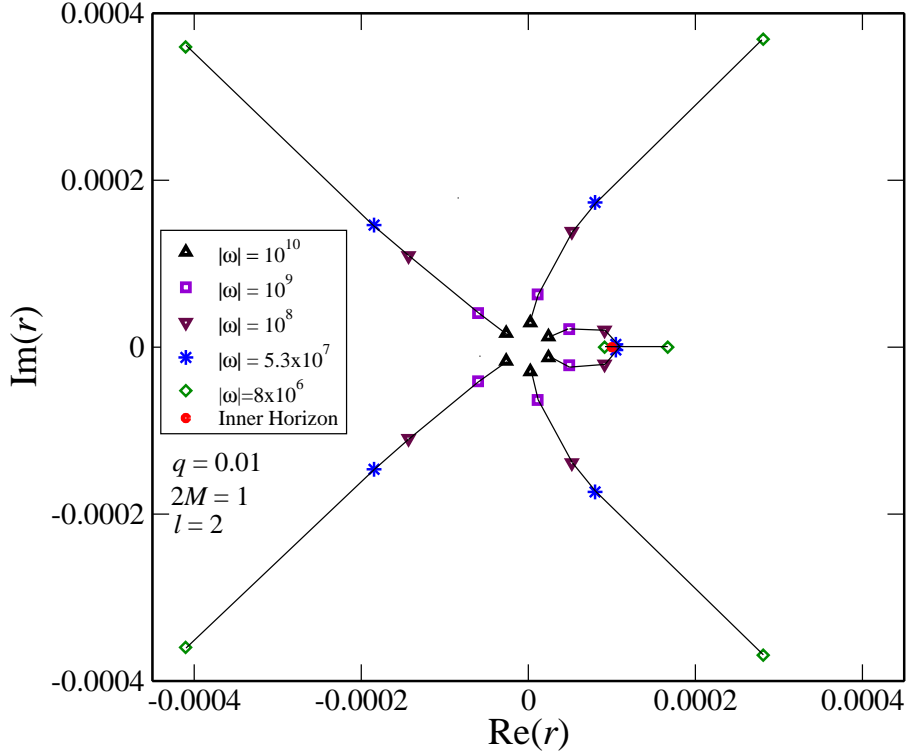}
\end{center}
\caption{The zeros of the function $Q$ for pure gravitational axial perturbations in four spacetime dimensions for different values of $|\omega|$.}
\label{zeros}
\end{figure}

\begin{figure}[tb]
\begin{center}
\includegraphics[height=4cm]{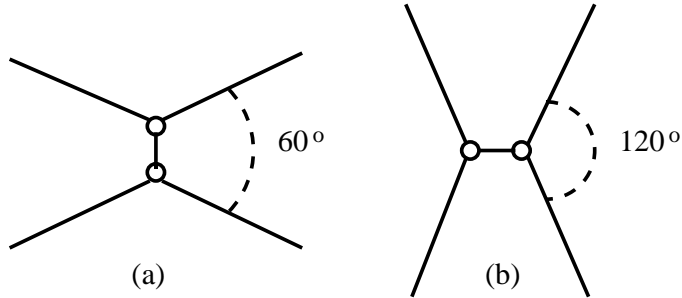}
\end{center}
\caption{Anti-Stokes lines emanating from the two zeros {\it{before}} (a) and {\it {after}} (b) they coincide on the real axis.}
\label{phasetrans}
\end{figure}

\begin{figure}[tb]
\begin{center}
\includegraphics[height=8cm]{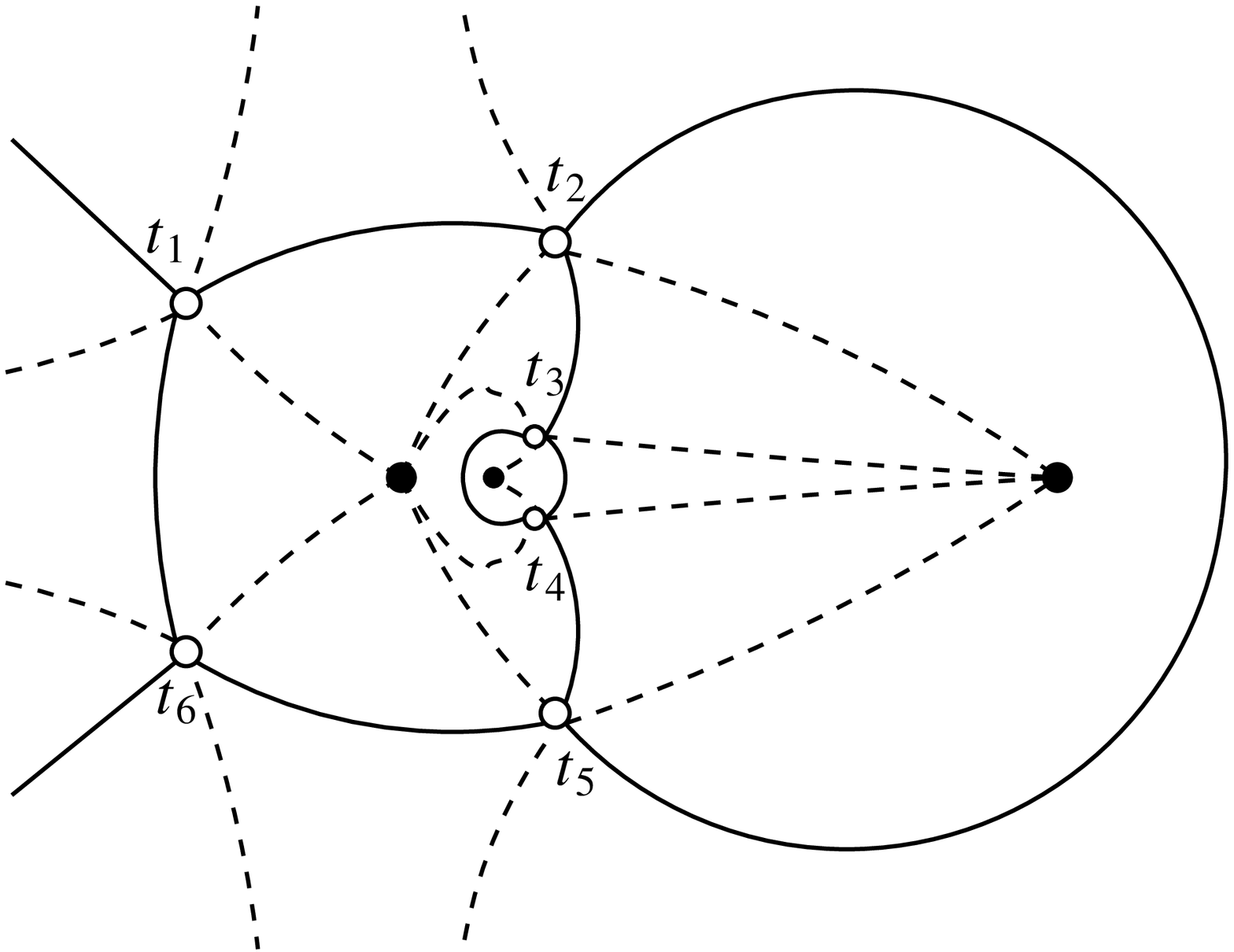}
\end{center}
\caption{Schematic presentation of Stokes (dashed) and Anti-Stokes (solid) lines in the complex plane {\it{before}} the two zeros coincide.  The open circles are the zeros and the filled circles are the poles of the function $Q^2$.}
\label{RN-schematic1}
\end{figure}

\begin{figure}[tb]
\begin{center}
\includegraphics[height=8cm]{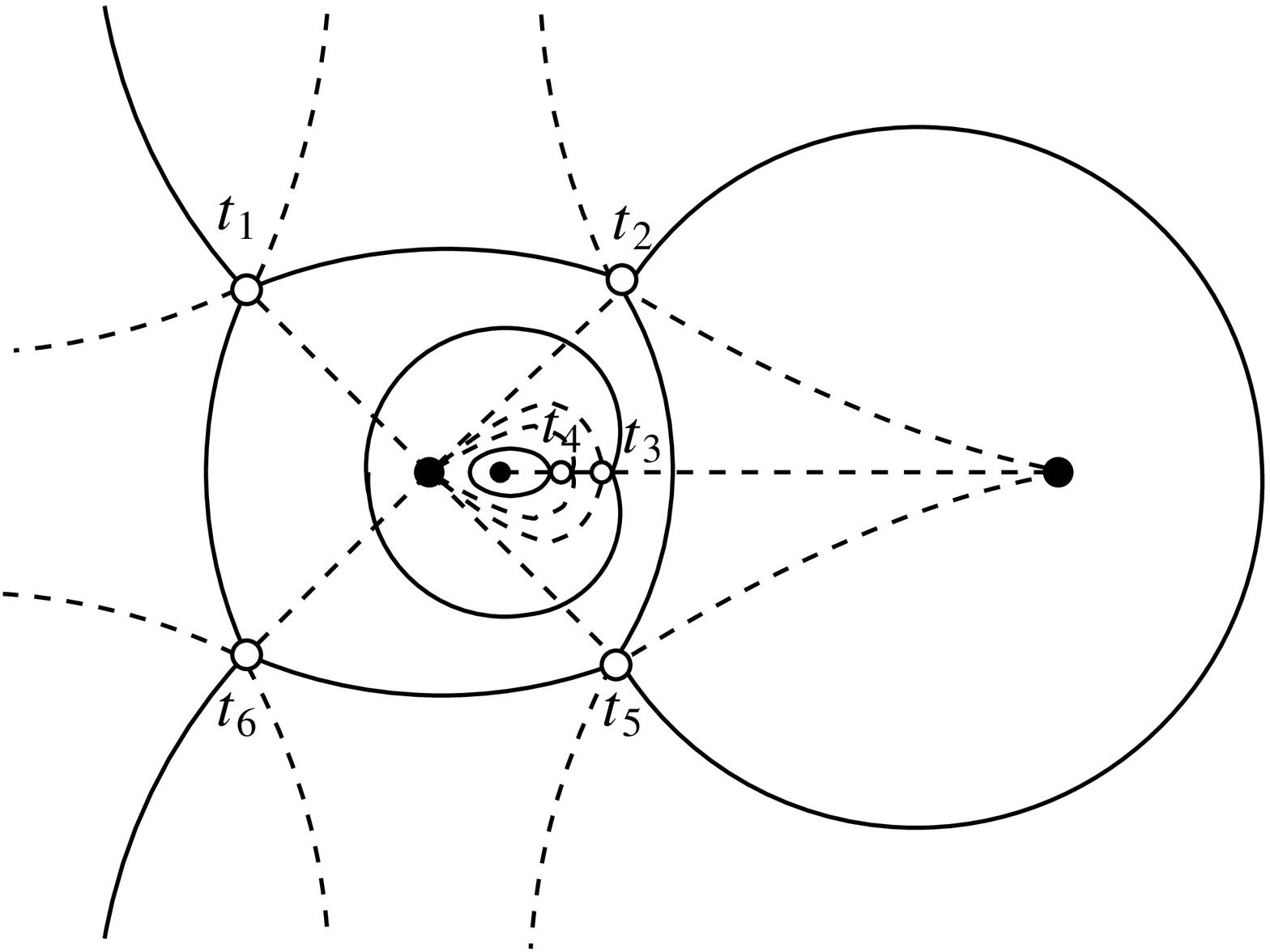}
\end{center}
\caption{Schematic presentation of Stokes (dashed) and Anti-Stokes (solid) lines in the complex plane {\it{after}} the two zeros coincide.  The open circles are the zeros and the filled circles are the poles of the function $Q^2$.}
\label{RN-schematic2}
\end{figure}

\begin{figure}[tb]
\begin{center}
\includegraphics[height=10cm]{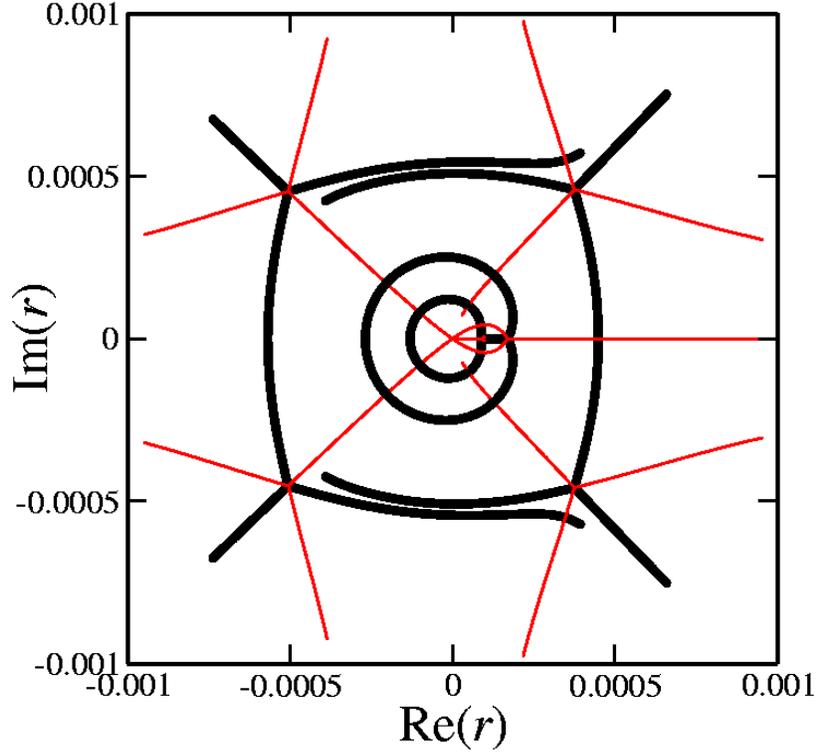}
\end{center}
\caption{Stokes (thin) and anti-Stokes (thick) lines produced by numerical calculations for pure gravitational axial perturbations, where $l=2$, $2M=1$, and $q=0.01$.}
\label{RN-numeric}
\end{figure}

\begin{figure}[tb]
\begin{center}
\includegraphics[height=10cm]{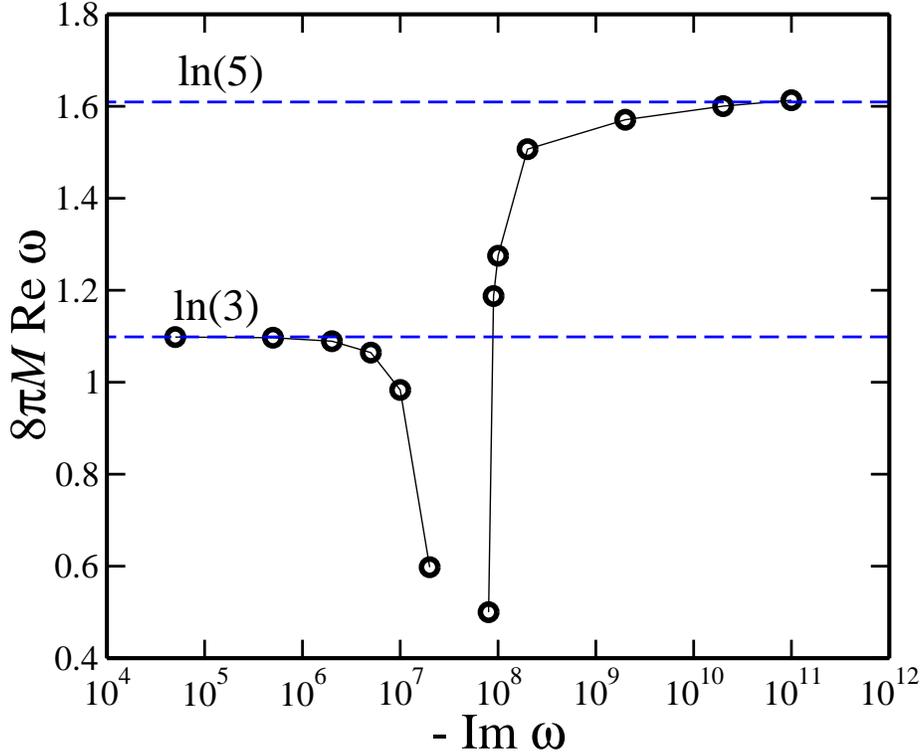}
\end{center}
\caption{$8\pi M {\mbox {Re}}~ \omega $ as a function of $-{\mbox {Im}}~ \omega$ for pure gravitational axial perturbations, where $l=2$, $2M=1$, and $q=0.01$.}
\label{RN-Sch-trans}
\end{figure}

\begin{figure}[tb]
\begin{center}
\includegraphics[height=10cm]{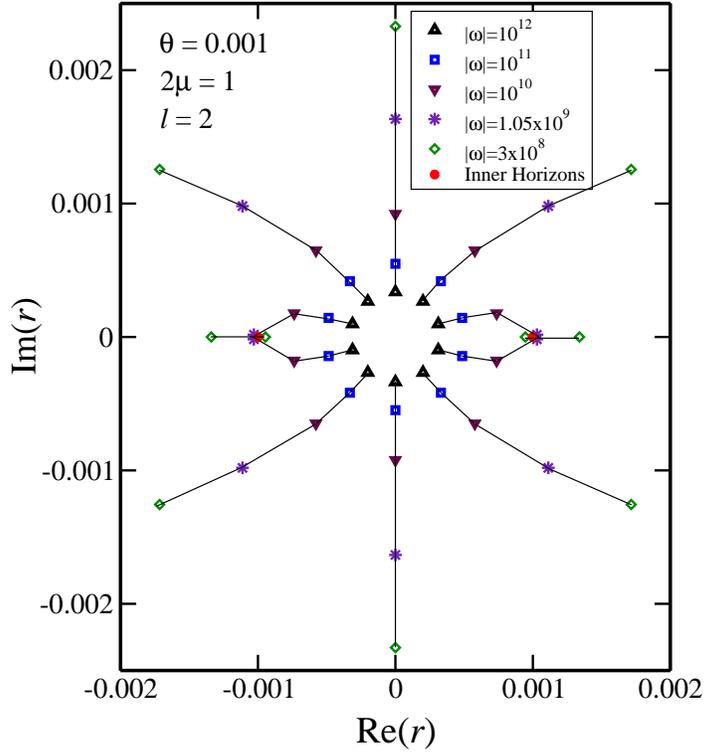}
\end{center}
\caption{The zeros of the function $Q$ for pure gravitational vector perturbations in 5-$d$ for different values of $|\omega|$.}
\label{zeros-5d}
\end{figure}

\begin{figure}[tb]
\begin{center}
\includegraphics[height=10cm]{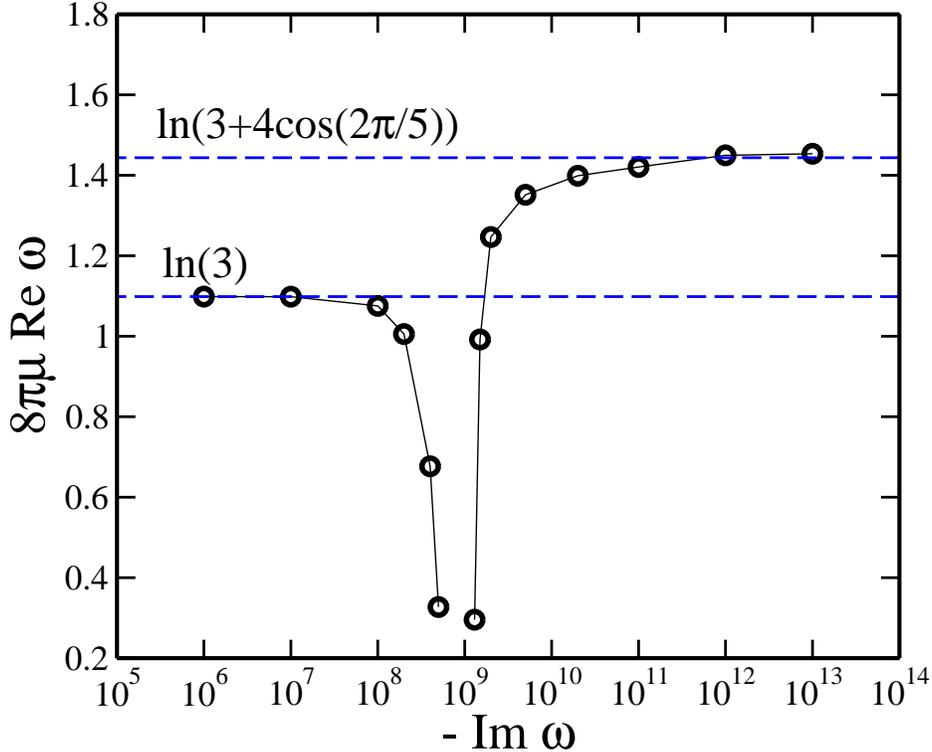}
\end{center}
\caption{$8\pi\mu {\mbox {Re}}~ \omega $ as a function of $-{\mbox {Im}}~ \omega$ for pure gravitational vector perturbations in 5-$d$, where $l=2$, $2\mu=1$, and $\theta=0.001$.}
\label{RN-Sch-trans-5d}
\end{figure}

\sxn{Schwarzschild Limit of Reissner-Nordstr$\ddot{\mbox{o}}$m Black Holes}

It was shown in the previous section that the infinitely damped QNM frequencies of R-N black holes do not approach the Schwarzschild result in the limit when the black hole charge goes to zero.  This is a potentially puzzling issue that deserves more careful investigation.  The resolution was suggested in \cite{Andersson}: there should be an intermediate regime of large but finite damping in which QNMs resemble the Schwarzschild result. We now investigate this issue using a combination of analytic and numerical methods to carefully track the modification in the structure of the Stokes and anti-Stokes lines in the complex plane as we approach the Schwarzschild limit.
We will of necessity concentrate only on vector perturbations because in tensor and scalar perturbations the constant $J$ is zero for Schwarzschild black holes where $\theta=0$, which makes the  numerical analysis problematic in the Schwarzschild limit. Note that the analytic calculations stay away from $J=0$ until the end, and then take the limit $J\to0$ in the final expression for the frequencies.

In the case of R-N black hole, for vector electromagnetic-gravitational perturbations one has two distinct effective potentials
\beeq
V_{v\pm}= f(r)\left[ -{n(n-2)\over 4}\lambda+{{4l(l+n-1)+n(n-2)}\over {4r^2}}+{{\beta_\pm}\mu\over{r^{n+1}}}+{{n(5n-2)\theta^2}\over{4r^{2n}}} \right]~,
\label{Vaxial}
\eneq 
where 
\beeq
\beta_{\pm}=-{n^2+2 \over 2} \pm \sqrt{(n^2-1)^2+2n(n-1)[l(l+n-1)-n]{\theta^2\over \mu^2}}~.
\label{Vaxial-beta}
\eneq 
Setting $\lambda=0$ we get
\begin{eqnarray}
R_{v\pm}(r)  & = & {1 \over f^2} \left \{ \omega^2  -{{4l(l+n-1)+n(n-2)}\over {4r^2}}+{{4l(l+n-1)+3n^2-4n-\beta_\pm}\over {2r^{n+1}}}\mu  \right. \nonumber \\
      &      &  \mbox{} ~~~~~ \left. -{n^2-1-2\beta_\pm \over r^{2n}}\mu^2-\frac{n(7n-5)+{n\over4}(n-2)+l(l+n-1)}{r^{2n}}\theta^2  \right. \nonumber \\
      &      &  \mbox{} ~~~~~ \left. +{3n(n-1)+{n\over 2}(5n-2)-\beta_\pm  \over {r^{3n-1}}}\mu\theta^2-\frac{n(n-1)+{n\over 4}(5n-2)}{r^{4n-2}}\theta^4 \right\}~.
\label{R-RN}
\end{eqnarray}
Since we are interested in the Schwarzschild limit, throughout this paper we take $\theta/\mu \ll 1$, where the coupled electromagnetic and gravitational vector perturbations approach pure electromagnetic and gravitational perturbations.  In this limit, for electromagnetic waves we have $\beta_+\approx n^2/2-2$ and for gravitational waves we have $\beta_-\approx -3n^2/2$.  In order to do numerical analysis, let us concentrate first on four dimensional spacetime where the function $R$ for pure electromagnetic and gravitational perturbations is
\beeq
R(r)   \approx {1 \over f^2} \left \{ \omega^2  -\frac {l(l+1)}{r^2}+\frac{j^2+l(l+1)}{r^3} 2M
  - \frac{j^2-1/4}{r^4}(2M)^2 
       +\frac{j^2+6}{r^5}2Mq^2-\frac{6}{r^6}q^4 \right\}~.
\label{R-RN-EM-Grav}
\eneq
Here $q$ and $M$ are the charge and mass of the black hole respectively and $j$ is the spin of the perturbation which is $1$ for electromagnetic and $2$ for gravitational perturbations.
Note that in the infinite damping limit, where $-{\mbox {Im}}~\omega \rightarrow \infty$, the zeros of the function $R$ approach the origin of the complex plane and as in Eq. (\ref{Rr2}) one can always take
\beeq
R(r)\sim \frac{\omega^2}{f^2}-\frac{6}{r^2}~.
\eneq
On the other hand, for large but finite damping, where $-{\mbox {Im}}~\omega \gg {\mbox {Re}}~\omega$, the zeros of the function $R$ are not infinitely close to the origin of the complex plane.  In this case we need in principle to consider all the terms in the function $R$.  
The terms in Eq. (\ref{R-RN-EM-Grav}) of order $r^{-3}$ and $r^{-2}$ inside the curly bracket are in fact negligible as long as the zeros of the function $R$ are located at a radius much smaller than $1$, but for the sake of accuracy, we nonetheless keep all the terms in Eq. (\ref{R-RN-EM-Grav}) in our numerical calculations. 

Fig. \ref{zeros} shows how the zeros of the function $Q^2=R-{1\over 4r^2}$, for pure gravitational axial perturbations ($j=2$), move in the complex plane as we change the value of $|\omega|$.  (In our calculations we always assume that $|\omega| = |{\mbox {Im}}~\omega |$.)
In this figure we have taken $l=2$, $2M=1$, and $q=0.01$.  As $|\omega|$ decreases, four of the zeros move away from the origin and approach the Schwarzschild limit in which the zeros are located at $r_0 \approx |r_0|e^{(2k+1) \pi/4}$, where $k=0, 1, 2, 3$.  The remaining two zeros, located on the right side of the origin, initially move away from the origin and then they move toward the positive real axis, where they eventually  coincide  at a point to the right of the inner horizon.  In Fig. \ref{zeros}, the two zeros coincide when the damping is approximately equal to $5.3\times 10^7$.
After they coalesce, one zero moves toward the inner horizon and the other one moves away from the inner horizon. This motion does not continue indefinitely. In Fig. \ref{zeros}, the two zeros on the real axis asymptote to $r=0.00009122\cdots$, which is just inside the inner horizon, and $r=0.00017133\cdots$, which is just outside the inner horizon.  Thus, two of the zeros stay relatively close to the inner horizon while the other zeros move far away from the origin.  These four zeros finally enter a region in the complex plane far enough away from the origin so that the terms with the charge $q$ in the functions $f$, $R$, and $Q$ become negligible.  This happens in the range
\beeq
\frac{(j^2-1)+3/4}{(2M)^2} \ll |\omega|^2 \ll \frac{4(j^2-1)+3}{M^2}\left(\frac{M}{q}\right)^8~.
\label{Sch-range}
\eneq
For $q/M \ll 1$, this nonetheless allows many orders of magnitude for $|{\rm Im}~\omega|$.  In this range, we can approximate $Q^2$ to be 
\beeq
Q^2(r) \sim \frac{1}{f^2}\left\{ \omega^2-\frac{[(j^2-1)+3/4](2M)^2}{r^4}\right\}-{1 \over 4r^2}~,
\eneq
which is exactly identical to what we have for Schwarzschild black holes. 

It is important to look carefully at what happens when two of the zeros merge on the real axis. Fig. {\ref{phasetrans}(a), shows the two zeros just before they coincide on the real axis.  The two anti-Stokes lines to the left  encircle the inner horizon and the two anti-Stokes lines to the right connect to the neighboring zeros on either side of the positive real axis.  After the zeros coincide the angle between the anti-Stokes lines on the right hand side, which was $60^o$ before they coincide, has changed to $120^o$ as one can see in Fig. {\ref{phasetrans}.  During this transition, the lines must break apart.  The anti-Stokes lines to the right  detach from their neighboring zeros and  encircle the pole at the origin.  The schematic behavior of the Stokes and anti-Stokes lines {\it{before}} and {\it after} the two zeros coincide on the real axis are shown in Figs. {\ref{RN-schematic1}} and {\ref{RN-schematic2}} respectively.  These schematic figures are based on numerically generated diagrams of Stokes and anti-Stokes lines in the complex plane.  An example of such numerically generated diagrams is shown in Fig. \ref{RN-numeric}, which plots the Stokes and anti-Stokes lines for the gravitational perturbations  ($j=2$) with $l=2$, $2M=1$, $q=0.01$, and $|\omega|=5 \times 10^6$.  Note that Fig \ref{RN-numeric} shows that at this point one of the zeros is located inside the inner horizon.  

The numerical recipe we have written is operating in the following way:  The input parameters are $j$, $M$, $q$, $\omega$, the exact position of the zeros of the function $Q$, and the approximate angles in which the Stokes and anti-Stokes lines emanate from those zeros.  In the program we use the fact that
\beeq
\int_{a}^{a+dz} \sqrt{\tilde{Q}(r)}dr = {2\over 3}{1\over \tilde{Q}'(a)}\left[(\tilde{Q}(a)+\tilde{Q}'(a)dz)^{3/2}-\tilde{Q}(a)^{3/2}\right]~,
\label{numeric1}
\eneq
when the stepsize $|dz|$ is small enough.  Here $\tilde{Q}=Q^2$.  The numerical calculation starts at the zeros of the function $Q$. To find the Stokes lines, the program finds the point where
\beeq
-\epsilon \leq {\rm Re} \left\{ {2\over 3}{1\over \tilde{Q}'(a)}\left[(\tilde{Q}(a)+\tilde{Q}'(a)dz)^{3/2}-\tilde{Q}(a)^{3/2}\right] \right\} \leq \epsilon~,
\label{numeric2}
\eneq 
and to find the anti-Stokes lines the program finds the points where
\beeq
-\epsilon \leq {\rm Im} \left\{ {2\over 3}{1\over \tilde{Q}'(a)}\left[(\tilde{Q}(a)+\tilde{Q}'(a)dz)^{3/2}-\tilde{Q}(a)^{3/2}\right] \right\} \leq \epsilon~.
\label{numeric3}
\eneq 
Here $\epsilon$ ideally needs to be zero but in the numerical calculations we take it to be a small number which generally works well if we take it to be roughly equal to the stepsize $|dz|$.   
As the program calculates the direction of the lines and moves away from the zeros the numerical error gets bigger.  This is the reason why some times the lines miss each other and they move close to each other until the program cannot find any more points as seen in Fig. \ref{RN-numeric}.  

We can use also the numerically generated Stokes and anti-Stokes lines to calculate the QNM frequency.  First, we use the method of {\cite{Andersson}} to find the general WKB condition
\beeq
e ^{2i\Gamma}=-e^{2i\gamma _{52}}-{(1+e ^{2i\gamma_{52}}) \over e ^{2i\gamma_{12}}}~
\label{WKB-Sch-generic}
\eneq 
for the topology shown in Fig. \ref{RN-schematic2}.  Here
\beeq
\gamma_{ab}=\int^{t_b}_{t_a} Q(r)dr~,
\label{gab}
\eneq 
and $\Gamma$ is given in Eq. (\ref{Gamma}) when $n=2$.
In the Schwarzschild topology, where $\gamma_{12}=\gamma_{52}=-\pi$, Eq. (\ref{WKB-Sch-generic}) reduces to the WKB condition we found in Eq. (\ref{Sch-WKB}) in four spacetime dimensions. The phase integrals $\gamma_{12}$ and $\gamma_{52}$ can be calculated numerically.  For the particular case in Fig. \ref{RN-numeric} we find $\gamma_{12}\sim -3.076\cdots$ radians and $\gamma_{52}\sim -3.447\cdots$ radians.  Therefore we get
\beeq
e ^{2i\Gamma}=e^{8\pi \omega M}\approx -2.547\cdots+i1.380\cdots~.
\label{WKB-numeric1}
\eneq 
In other words $ {\mbox {Re}}~ \omega M \approx \ln(2.897 \cdots )/8 \pi$.

Similar to Eq. (\ref{WKB-Sch-generic}), we can find a general expression for the WKB condition for the topology shown in Fig. \ref{RN-schematic1}:
\beeq
e ^{2i\Gamma_e}=1-(1+e^{2i\gamma _{12}})(1+e^{-2i\gamma _{32}})e^{-2i\gamma _{12}}e^{2i\gamma _{32}}(1+e^{-2i\Gamma_i}e^{2i\tilde \gamma_{43}}e^{2i\gamma _{32}})~,
\label{WKB-RN}
\eneq 
where $\Gamma_e$ and $\Gamma_i$ are given in Eqs. (\ref{Gamma-e}) and (\ref{Gamma-i}) respectively when $n=2$.  To obtain the WKB condition (\ref{WKB-RN}), we have used the fact that $\gamma_{32}=\gamma_{54}$ by symmetry.  Note that the WKB conditions (\ref{WKB-Sch-generic}) and (\ref{WKB-RN}) are also valid in higher spacetime dimensions.

In the limit $q/M \ll 1$, we can write
\beeq
\Gamma_e \approx -4 \pi i \omega M \left ( 1+{q^4 \over 16M^4} \right ) \approx -4 \pi i \omega M ~,
\label{Gamma_e}
\eneq 
and
\beeq
\Gamma_i \approx \pi i \omega M \left( {q^4 \over 4M^4} \right )~.
\label{Gamma_i}
\eneq 
Therefore, Eq. (\ref{WKB-RN}) can be approximated to be
\beeq
e ^{8\pi \omega M} \approx 1-(1+e^{2i\gamma _{12}})(1+e^{-2i\gamma _{32}})e^{-2i\gamma _{12}}e^{2i\gamma _{32}}(1+e^{2i \gamma_{43}}e^{2i\gamma _{32}})~.
\label{WKB-RN-approx}
\eneq 
We can now numerically calculate the phase integrals $\gamma_{12}$, $\gamma_{32}$, and $\gamma_{43}$ and evaluate the real part of $\omega$ by using the WKB condition (\ref{WKB-RN-approx}).

Fig. \ref{RN-Sch-trans} plots $8\pi M {\mbox {Re}}~ \omega$ as a function of $-{\mbox {Im}}~ \omega$ for pure gravitational axial perturbations, where we have taken $l=2$, $2M=1$, and $q=0.01$. It is interesting to note that at the point where the topology changes from  R-N to Schwarzschild, the real part of the QNM frequency approaches to zero.

\sxn{Schwarzschild Limit in Higher Spacetime Dimensions}

So far we have concentrated on the metric perturbations in four spacetime dimensions.  The same type of numerical analysis can readily be done in higher spacetime dimensions and one expects similar results.  The crucial point is that as the magnitude of the frequency $\omega$ changes, there should be a topology change of the kind which is shown in Figs. \ref{RN-schematic1} and \ref{RN-schematic2}.  This type of topology change happens when two of the zeros of the function $Q$, which are off the real axis initially, coincide on the real axis and then move along the real axis staying relatively close to the inner Cauchy horizon. We plot the zeros of the function $Q$ for pure gravitational vector perturbations in five spacetime dimensions for different values of $|\omega|$ in Fig. \ref{zeros-5d} and give the corresponding results for the real part of the QNM frequencies in Fig. \ref{RN-Sch-trans-5d} respectively.

We expect this behaviour is generic to all dimensions: in $d=n+2$ dimensions there are, in the R-N case, two times
$(n-1)$ horizons distributed in pairs along equally spaced radial lines in the complex plane. The physical pair is, of course, on the real axis. In the Schwarzschild limit $Q^2$ has $2n$ zeros, while in the R-N limit
it has $4n-2$. The difference is $2(n-1)$ zeros, which converge in pairs near the $n-1$ inner horizons  and then move
apart at right angles. This behaviour should occur in the range:
\beeq
1 \ll |\omega|^2 \mu^{2/(n-1)}\ll \left(\frac{\mu}{\theta}\right)^{4n/(n-1)}~.
\label{Sch-range1}
\eneq
 There will be a  critical value of the damping in this range for which the pairs coincide and the real frequency will be zero or near zero.  In Figs. \ref{zeros-5d} and \ref{RN-Sch-trans-5d}, this critical damping happens at approximately $1.05\times 10^9$.

\sxn{ Conclusions}

We have calculated explicitly the QNM frequencies of R-N black holes in the transition region of damping that interpolates between the R-N and Schwarzschild limits. Our numerical results suggest that the real part of the frequency decreases to zero from its $\ln(3)$ Schwarzschild value at a critical value of the damping before increasing to the R-N value. This behaviour is due to the discontinuous nature of the transition from six evenly spaced zeros of the WKB phase to four evenly spaced zeros, with two zeros even closer to the origin. In particular the two zeros first meet near the horizon at the critical value of the damping and then emerge at right angles. It is at this stage that the Stokes lines undergo a discrete topology change and the real part of the frequency goes to zero. We have shown that this behaviour is also present in five dimensions, and argued that it is a generic feature of all higher dimensional R-N black holes.

The physical interpretation of the highly damped QNM frequencies in terms of the quantum spectrum of black holes is highly speculative, and seems to be contradicted by a myriad of results for multi-horizon, non-asymptotically flat black holes. It should be noted, however, that Hod's original conjectures only apply directly to single horizon black holes, whose horizons are described by one dimensionful parameter, which could without loss of generality be taken as the temperature. The present analysis, in our opinion, seems to reverse slightly this negative trend. In effect, we have shown explicitly that for R-N black holes with small charge, the infinitely damped QNMs essentially ``see'' the full structure of the solution, including the inner horizon, which affects the value of the frequency in this limit. There is nonetheless an intermediate damping for which the QNMs only probe the outer horizon, and for this range the ``universal'' value of $\ln(3)$ is reproduced. 

Although our combination of analytic and numerical techniques does not allow us to determine whether 
the real part of the QNM frequency goes exactly to zero, it is tempting to speculate that there is a 
connection between the value of the damping for which this happens and the algebraically special 
frequencies found by Chandrasekhar\cite{Chandra}. In the limit of small charge $q$, 
one of these values (the one associated with the almost pure electromagnetic perturbations) 
occurs at $-i[3l(l+1)M/(4 q^2)]$,
which is the same order of magnitude for which our special frequency occurs.  However, our frequency 
appears to be independent of $l$.  Moreover, we have examined in detail only gravitational 
axial perturbations, for which Chandrasekhar's algebraically special frequency gets very small. 
Nonetheless it is important to discover the precise nature of the special frequency that occurs in 
our analysis, and its connection with other algebraically special frequencies. This is currently 
under investigation.

Hod's physical interpretation of the highly damped QNM frequencies in terms of the quantum spectrum 
of black holes is highly speculative, and seems to be contradicted by a myriad of results for multi-horizon, 
non-asymptotically flat black holes.  It should be noted, however, that an equally spaced area spectrum for 
black holes has been derived by several authors using a variety of techniques\cite{equal area}.  Moreover, 
Hod's original conjecture only applies directly to single horizon black holes, whose horizons are described 
by one dimensionful parameter, which could without loss of generality be taken as the temperature.\footnote{An 
intriguing speculation by Makela {\it et. al.}\cite{Makela} may be relevant in this regard: they propose 
that for multi-horizon black holes it is the total area of all horizons that has an equally spaced spectrum.} 
In any case, the present analysis, in our opinion, seems to reverse slightly the current trend, and lends 
support to the possible validity of Hod's conjecture.  In effect, we have shown explicitly that for R-N black 
holes with small charge, the infinitely damped QNMs essentially ``see'' the full structure of the solution, 
including the inner horizon, which affects the value of the frequency in this limit. There is nonetheless 
an intermediate damping for which the QNMs only probe the outer horizon, and for this range the ``universal'' 
value of $\ln(3)$ is reproduced. 
It therefore seems that this issue is worthy of continued investigation.

\vskip .5cm

\leftline{\bf Acknowledgments}
This research was supported in part by the Natural Sciences and Engineering Research Council of Canada.

%%%%%%%%%%%%%%%%%%%%%%%%%%%%%%

% A useful Journal macro
\def\jnl#1#2#3#4{{#1}{\bf #2} (#4) #3}

\def\Zphys{{\em Z.\ Phys.} }
\def\jssc{{\em J.\ Solid State Chem.\ }}
\def\jpsJ{{\em J.\ Phys.\ Soc.\ Japan }}
\def\ptps{{\em Prog.\ Theoret.\ Phys.\ Suppl.\ }}
\def\PTP{{\em Prog.\ Theoret.\ Phys.\  }}
\def\LNC{{\em Lett.\ Nuovo.\ Cim.\  }}

\def\JMP{{\em J. Math.\ Phys.} }
\def\NPB{{\em Nucl.\ Phys.} B}
\def\NP{{\em Nucl.\ Phys.} }
\def\PLB{{\em Phys.\ Lett.} B}
\def\PL{{\em Phys.\ Lett.} }
\def\PRL{\em Phys.\ Rev.\ Lett. }
\def\PRB{{\em Phys.\ Rev.} B}
\def\PRD{{\em Phys.\ Rev.} D}
\def\PR{{\em Phys.\ Rev.} }
\def\PRe{{\em Phys.\ Rep.} }
\def\AP{{\em Ann.\ Phys.\ (N.Y.)} }
\def\RMP{{\em Rev.\ Mod.\ Phys.} }
\def\ZPC{{\em Z.\ Phys.} C}
\def\SCI{\em Science}
\def\CMP{\em Comm.\ Math.\ Phys. }
\def\MPLA{{\em Mod.\ Phys.\ Lett.} A}
\def\IJMPB{{\em Int.\ J.\ Mod.\ Phys.} B}
\def\cmp{{\em Com.\ Math.\ Phys.}}
\def\JPA{{\em J.\  Phys.} A}
\def\CQG{\em Class.\ Quant.\ Grav.~}
\def\ATMP{\em Adv.\ Theoret.\ Math.\ Phys.~}
\def\PRSA{{\em Proc.\ Roy.\ Soc.\ Lond.} A }
\def\ibid{{\em ibid.} }
\vskip 1cm

\leftline{\bf References}

\renewenvironment{thebibliography}[1]
        {\begin{list}{[$\,$\arabic{enumi}$\,$]}  
% {\arabic{enumi}.}
        {\usecounter{enumi}\setlength{\parsep}{0pt}
         \setlength{\itemsep}{0pt}  \renewcommand{\baselinestretch}{1.2}
         \settowidth
        {\labelwidth}{#1 ~ ~}\sloppy}}{\end{list}}

%%%%%%%%%%%%%%%%%%

\end{document}